%% file: article_vuillot_fault_tolerant_demo.tex
\documentclass[twoside]{article}
\usepackage{qic}

\usepackage[utf8]{inputenc}
\usepackage[english]{babel}
\usepackage[T1]{fontenc}
\usepackage{amsmath}
\usepackage[hidelinks]{hyperref}
\usepackage{qcircuit}
\usepackage{braket}
\usepackage{dsfont}
\usepackage{multirow}
\usepackage{tikz,pgfplots}
\usepackage{pgfplotstable}
\usepgfplotslibrary{patchplots}
\pgfplotsset{compat=1.7}
\usepackage{booktabs}
\usepackage{standalone}
\usetikzlibrary{arrows, decorations.markings,fit}
\usepackage{siunitx}
\usepackage{xcolor}
\usepackage[numbers,sort&compress]{natbib}

\tikzstyle{vecArrow} = [thick, decoration={markings,mark=at position
	1 with {\arrow[semithick]{open triangle 60}}},
double distance=1.4pt, shorten >= 5.5pt,
preaction = {decorate},
postaction = {draw,line width=1.4pt, white,shorten >= 4.5pt}]

\newcommand{\id}{\mathds{1}}
\newcommand{\co}[1]{\color{#1}}

\definecolor{color1}{RGB}{190,139,57}
\definecolor{color2}{RGB}{124,115,203}
\definecolor{color3}{RGB}{118,167,76}
\definecolor{color4}{RGB}{196,90,156}
\definecolor{color5}{RGB}{70,193,154}
\definecolor{color6}{RGB}{199,85,81}

\begin{document}
\setlength{\textheight}{8.0truein}    

\runninghead{Is Error Detection Helpful on IBM~5Q Chips ?}
            {Christophe Vuillot}

\normalsize\textlineskip
\thispagestyle{empty}
\setcounter{page}{1}

\copyrightheading{0}{0}{0000}{000--000}

\vspace*{0.88truein}

\alphfootnote

\fpage{1}

\centerline{\bf
IS ERROR DETECTION HELPFUL ON IBM~5Q CHIPS ?}
\vspace*{0.37truein}
\centerline{\footnotesize
CHRISTOPHE VUILLOT\footnote{c.vuillot@tudelft.nl}}
\vspace*{0.015truein}
\centerline{\footnotesize\it 1 - JARA Institute for Quantum Information, RWTH Aachen University}
\baselineskip=10pt
\centerline{\footnotesize\it Physikzentrum Campus Melaten
	Otto-Blumenthal-Strasse,
	52074 Aachen, Germany}
\vspace*{0.015truein}
\centerline{\footnotesize\it 2 - QuTech, TU Delft}
\baselineskip=10pt
\centerline{\footnotesize\it Lorentzweg 1, 2628 CJ Delft, Netherlands}
\vspace*{0.225truein}
\publisher{(received date)}{(revised date)}

\vspace*{0.21truein}

\abstracts{
This paper reports on experiments realized on several IBM~5Q chips which show evidence for the advantage of using error detection and fault-tolerant design of quantum circuits.
We show an average improvement of the task of sampling from states that can be fault-tolerantly prepared in the $[[4,2,2]]$ code, when using a fault-tolerant technique well suited to the layout of the chip.
By showing that fault-tolerant quantum computation is already within our reach, the author hopes to encourage this approach.
}{}{}

\vspace*{10pt}

\keywords{quantum, fault-tolerance, error detection, experiment}
\vspace*{3pt}
\communicate{to be filled by the Editorial}

\vspace*{1pt}\textlineskip    
\section{Introduction}

Quantum systems in laboratories around the world are reaching unprecedented level of control and precision for a variety of devices aiming at implementing reliable qubits. 
Yet, due to the very nature of those devices, errors are still notably present.
In order to be able to execute long quantum algorithms, improvements should be made using quantum error correction and fault-tolerant schemes.
So-called threshold theorems \cite{Aharonov1997,Knill365} prove that there exist error rates below which this approach is guaranteed to improve the device performance.
However, between those theorems and real experiments there remains a fog of technical details which clouds the actual practicality of error correcting codes.

Already some experiments which demonstrate the usefulness of quantum error correcting codes to protect a quantum memory have been done, e.g. \cite{Knill2001,Nigg2014,Corcoles2014,Linke2016,Takita2017}, but a demonstration of protected computation is still missing.

Inspired by a recent proposal by D. Gottesman \cite{Gottesman2016}, we use an IBM~5Q chip to show that error detection can improve some simple sampling tasks, thus turning a tiny portion of fog into blue sky.
Closely related to this work, \cite{Linke2016} and \cite{Takita2017} present experiments based on the same error detecting code, on trapped-ion qubits and on a similar superconducting chip, respectively.
Both those works focus on the preparation of few states and only one of the two encoded qubits is handled fault-tolerantly.
In this work an extensive set of fault-tolerant circuits is studied and both logical qubits are protected.
Moreover \cite{Linke2016} and \cite{Takita2017} introduce artificial Pauli errors to study the robustness of their preparation whereas this work takes a higher level approach, treating the experimental set-up as a black box, probing only the intrinsic errors in the system.
These considerations make the present work closer to the spirit of \cite{Gottesman2016}.
This work is, to the author's knowledge, the first experimental demonstration of error detection and fault-tolerance improving a quantum computation.

The paper is organized as follows. In \autoref{sec:principle} we present the principle of the approach and its specialization to the IBM~5Q chip. In \autoref{sec:results} the experimental results are shown and analyzed.

\section{Demonstrating fault-tolerance}
\label{sec:principle}

The idea behind fault-tolerance is to devise error correcting codes and the corresponding processes of encoding, correcting, computing or measuring with encoded information such that more errors are corrected than introduced \cite{Shor1996}.
The difficulty in devising such processes is that they are built out of components that are all faulty, so adding some might do more harm than it can help.
Devising, proving and simulating fault-tolerant schemes has been pursued for the last 20 years, e.g. \cite{Shor1996,Aharonov1997,Aliferis2006,Aliferis2008,Cross2009,Paetznick2011,Tomita2014,Chamberland2016}, but improving and finding better schemes is still important and a subject of ongoing research, e.g. \cite{Chao2017a,Chao2017,Takagi2017,Chamberland2017,Chamberland2017a}.

Taking one of these schemes and experimentally demonstrating its fault-tolerance, i.e finding an improvement of performance going from bare to encoded implementation, provides the ultimate validation of the scheme.

\subsection{General approach}

A description of a general approach to demonstrate fault-tolerant quantum computation is given in \cite{Gottesman2016}; we just briefly recall it here.
The idea is to choose a quantum error correcting code $\mathcal{C}$, admitting fault-tolerant circuits for the preparation of some logical states, denoted as $\ket{s_1},\dots,\ket{s_m}\in S_{\rm FT}^{\mathcal{C}}$, as well as for some logical quantum gates, denoted as $U_1,\dots,U_n\in G^{\mathcal{C}}_{\rm FT}$.
Using these as building blocks one can then randomly draw a state preparation from $S_{\rm FT}^{\mathcal{C}}$, then draw a sequence of gates from $G_{\rm FT}^{\mathcal{C}}$ to obtain an encoded fault-tolerant circuit.

More precisely, following the formalism of \emph{rectangles} and \emph{extended rectangle} introduced in \cite{Aliferis2006}, one interleaves the logical units (gates in $G^{\mathcal{C}}_{\rm FT}$, or preparation of states in $S^{\mathcal{C}}_{\rm FT}$) with a fault-tolerant circuit for error correction.
\emph{Rectangles} designate circuits comprising one logical unit preceded by a round of error correction, \emph{extended rectangles} designate circuits comprising one logical unit preceded and followed by a round of error correction.
The precise conditions to satisfy in order to be fault-tolerant for distance 3 codes are stated in \cite{Aliferis2006}. 
This ensures that the whole computation is error free if there is at most one fault per extended rectangle.

The error model considered in this paper is that the failure of any component can introduce any Pauli error on the qubits acted on by the component.
In our case, we cannot repeat fault-tolerant detection of errors, so our circuits will only tolerate a single fault in total.

The encoded circuits then have to be compared to a bare implementation on the physical qubits.
Sampling from the final state produced by the circuit in the computational basis and comparing the probability distribution obtained with the expected one gives a simple comparison metric.
Note that this supposes that the circuits are small enough or simple enough so that one can classically simulate sampling from the final state efficiently.
A successful demonstration of fault-tolerant quantum computation happens if the encoded circuits show better performance than the bare circuits.
To be the most convincing, one needs to use the best of the physical qubits as well as the most efficient implementation for the bare circuits.

\subsection{The IBM~5Q chip and [[4,2,2]]}

\begin{figure}[h]
	\centering
		\begin{minipage}{.45\textwidth}
		\centering
		{(a) Layout}\\
		\begin{tikzpicture}[node distance = 4.5em]
		\draw
		node[draw, circle,label = above:{\co{red}4}] (q2) {$q_2$}
		node[draw, circle, below right of = q2, label = right:{\co{red}1}] (q1) {$q_1$}
		node[draw, circle, below left of = q2, label = left:{\co{red}2}] (q3) {$q_3$}
		node[draw, circle, above left of = q2, label = left:{\co{red}3}] (q4) {$q_4$}
		node[draw, circle, above right of = q2, label = right:{\co{red}A}] (q0) {$q_0$};
		
		\path [draw, vecArrow] (q2) -- (q0)
		node[pos=0.5, below] (p) {};
		
		\path [draw, vecArrow] (q2) -- (q1)
		node[pos=0.5, below] (q) {};
		
		\path [draw, vecArrow] (q3) -- (q2)
		node[pos=0.5, below] (r) {};
		
		\path [draw, vecArrow] (q2) -- (q4)
		node[pos=0.5, below] (s) {};
		
		\path [draw, vecArrow] (q3) -- (q4)
		node[pos=0.5, below] (r) {};
		
		\path [draw, vecArrow] (q1) -- (q0)
		node[pos=0.5, below] (s) {};
		\end{tikzpicture}
	\end{minipage}
	 \begin{minipage}{.45\textwidth}
	 	\centering
		 {(b) $[[4,2,2]]$ Code}\\
		\begin{tikzpicture}[node distance = 2.7em]
		\draw
		node[draw, circle] (1) {\co{red}1}
		node[draw, circle, right of = 1] (2) {\co{red}2}
		node[draw, circle, below of = 1] (3) {\co{red}3}
		node[draw, circle, right of = 3] (4) {\co{red}4};
		\node[draw,ultra thick,rounded corners,inner sep = 1em, fit=(1) (2) (3) (4),label={[label distance=0em]right:{$S_X,S_Z$}}] {};
		\node[draw,green!70!black,ultra thick, inner sep = 0.2em, fit=(1) (3), rounded corners, label={[label distance=0.6em]below:{\co{green!70!black}$Z_2,X_1$}}] {};
		\node[draw,blue,ultra thick,rounded corners, inner sep = 0.6em, fit=(1) (2), label={[label distance=0.2em]above:{\co{blue}$Z_1,X_2$}}] {};
		\end{tikzpicture}
	 \end{minipage}
	 \fcaption{(a) The connectivity between the qubits in the IBM~5Q Raven chip. Arrows indicate CNOT capabilities where the arrow points towards the target of the CNOT. Red labels indicate our choice of qubit numbering. $A$ is an ancillary qubit used in verification of state preparation. (b) The usual representation layout for the $[[4,2,2]]$ code. The numbered circles represent the qubits and the rectangular boxes represent the support of different Pauli operators: logical Pauli operators $Z_1$, $X_2$ are highlighted in blue, logical Pauli operators $Z_2$, $X_1$ in green and stabilizers $S_X$, $S_Z$ in black.}
	\label{fig:chip}
\end{figure}

In 2016 IBM released a quantum chip with fixed-frequency superconducting transmon qubits, named IBM~5Q.
They provide worldwide cloud access to the chip under an initiative called the ``IBM Quantum Experience'' \cite{IBMQE}. 
The current iteration on which the experiments were done is nicknamed \emph{Raven}. The appendix also presents and compares preliminary results obtained with a previous iteration, \emph{Sparrow}.
The chip has five qubits, natively named $q_0$ to $q_4$.
It features single qubit Clifford gates, the $T$ gate ($\mathrm{diag}(1,\rm e^{i\pi/4})$) as well as some two-qubits CNOTs with a certain layout represented in Fig.~\ref{fig:chip}a.
This many qubits is the right number to use for a demonstration using the $[[4,2,2]]$ code as discussed in \cite{Gottesman2016}.

The $[[4,2,2]]$ code encodes two qubits into four physical qubits.
Its code space is stabilized by the all-$X$ and all-$Z$ Pauli operators ($S_X = X\otimes X\otimes X\otimes X$, $S_Z = Z\otimes Z\otimes Z\otimes Z$), together with the logical Pauli operators, they are represented in Fig.~\ref{fig:chip}b.
The logical code states are
\begin{eqnarray}
\ket{00}_{\rm L} &= \frac{1}{\sqrt{2}}\left (\ket{0000}+\ket{1111}\right )\label{eq:00}\\
\ket{01}_{\rm L} &= \frac{1}{\sqrt{2}}\left (\ket{1100}+\ket{0011}\right )\label{eq:01}\\
\ket{10}_{\rm L} &= \frac{1}{\sqrt{2}}\left (\ket{1010}+\ket{0101}\right )\label{eq:10}\\
\ket{11}_{\rm L} &= \frac{1}{\sqrt{2}}\left (\ket{1001}+\ket{0110}\right ).\label{eq:11}
\end{eqnarray}
The code also admits transversal implementations of two Clifford gates, namely $H\otimes H\cdot\mathrm{SWAP}$, where $H$ is the Hadamard gate, see Fig.~\ref{fig:gates}a, and the controlled phase or C$Z$ gate, see Fig.~\ref{fig:gates}b.
\begin{figure}[h]
	 \begin{minipage}{.5\textwidth}
	 	\centering
		 {(a)}
		\[
		\Qcircuit @C=.5em @R=.5em{
			&        &          &   &&                           &&&&&\lstick{\mathrm{q_1}}&\gate{H}&\qw\\
			\lstick{}&\gate{H}&\qswap    &\qw&&\raisebox{-2.5em}{$\equiv$}&&&&&\lstick{\mathrm{q_2}}&\gate{H}&\qw\\
			\lstick{}&\gate{H}&\qswap\qwx&\qw&&                           &&&&&\lstick{\mathrm{q_3}}&\gate{H}&\qw\\
			&        &          &   &&                           &&&&&\lstick{\mathrm{q_4}}&\gate{H}&\qw
		}\]
	 \end{minipage}
	 \begin{minipage}{.5\textwidth}
	 	\centering
		{(b)}
		\[\Qcircuit @C=.5em @R=.5em{
			&        &        &        &   &&                           &&&&&\lstick{\mathrm{q_1}}&\gate{S}&\qw\\
			\lstick{}&\qw     &\ctrl{1}&\qw     &\qw&&\raisebox{-2.5em}{$\equiv$}&&&&&\lstick{\mathrm{q_2}}&\gate{S}&\qw\\
			\lstick{}&\gate{H}&\targ   &\gate{H}&\qw&&                           &&&&&\lstick{\mathrm{q_3}}&\gate{S}&\qw\\
			&        &        &        &   &&                           &&&&&\lstick{\mathrm{q_4}}&\gate{S}&\qw
		}\]
	 \end{minipage}
	 \fcaption{(a) Bare and encoded versions of $H\otimes H\cdot\mathrm{SWAP}$, where $H$ is the Hadamard gate. In practice the SWAP gate is not done physically on the bare version but just in software, by renumbering the two qubits. (b) Bare and encoded versions of the $\mathrm{C}Z$ gate, where $S=\mathrm{diag}(1,i)$.}\label{fig:gates}
\end{figure}
Moreover, if the five qubits have the right connectivity then there are fault-tolerant circuits preparing the logical states $\ket{00}_{\rm L}$, $\ket{0+}_{\rm L}$ and $\ket{00}_{\rm L}+\ket{11}_{\rm L}$.
For the cat state $\ket{00}_{\rm L}$, see Equation (\ref{eq:00}), there exists fault-tolerant preparations including one ancillary qubit to verify the preparation.
One post-selects on a successful preparation.
The logical states $\ket{0+}_{\rm L}$ and $\ket{00}_{\rm L}+\ket{11}_{\rm L}$ are both two Bell pairs but with different pairings.
They both also have fault-tolerant preparation circuits.
In \cite{Gottesman2016}, some circuits are proposed, they can be adapted to the layout of the chip. We tested another preparation for $\ket{00}_\mathrm{L}$ inspired by \cite{Chao2017a}, which is substantially shorter so more relevant to this layout.

\begin{figure}[h]
	\centering
	 \begin{minipage}{.4\textwidth}
	 	\centering
	 	{(a) FTv1}
		\[
		\Qcircuit @C=.5em @R=.5em{
			\lstick{\mathrm{q_0}\rightarrow\ket{0}}&\qw     &\targ    &\qw      &\qw     &\qw      &\qw     &\qw     &\targ    &\meter&\cw\\ 
			\lstick{\mathrm{q_1}\rightarrow\ket{0}}&\qw     &\qw      &\targ    &\qw     &\qw      &\qw     &\qw     &\qw      &\qw   &\qw\\
			\lstick{\mathrm{q_2}\rightarrow\ket{0}}&\gate{H}&\ctrl{-2}&\ctrl{-1}&\gate{H}&\targ    &\gate{H}&\ctrl{2}&\ctrl{-2}&\qw   &\qw\\
			\lstick{\mathrm{q_3}\rightarrow\ket{0}}&\qw     &\qw      &\qw      &\gate{H}&\ctrl{-1}&\gate{H}&\qw     &\qw      &\qw   &\qw\\
			\lstick{\mathrm{q_4}\rightarrow\ket{0}}&\qw     &\qw      &\qw      &\qw     &\qw      &\qw     &\targ   &\qw      &\qw   &\qw 
		}\]
	 \end{minipage}\qquad\qquad
	 \begin{minipage}{.4\textwidth}
	 	\centering
		 {(b) FTv2}
		\[
		\Qcircuit @C=.5em @R=.5em{
			\lstick{\ket{0}}&\qw     &\qw      &\qw     &\qw      &\qw       &\targ    &\targ    &\meter&\cw\\ 
			\lstick{\ket{0}}&\qw     &\qw      &\qw     &\targ    &\qw       &\qw      &\ctrl{-1}&\qw   &\qw\\
			\lstick{\ket{0}}&\qw     &\targ    &\gate{H}&\ctrl{-1}&\qswap    &\ctrl{-2}&\qw      &\qw   &\qw\\
			\lstick{\ket{0}}&\gate{H}&\ctrl{-1}&\gate{H}&\ctrl{1} &\qw\qwx   &\qw      &\qw      &\qw   &\qw\\
			\lstick{\ket{0}}&\qw     &\qw      &\qw     &\targ    &\qswap\qwx&\qw      &\qw      &\qw   &\qw
		}\]
	 \end{minipage}
	 \fcaption{The two different implementations for preparing the state $\ket{00}_\mathrm{L}$: (a) Based on a flagging technique with 5 CNOTs, (b) based on the circular connectivity circuit with 8 CNOTs.
		In both circuits if the ancillary qubit $\mathrm{q}_0$ is measured as 1, we reject the preparation.}
	\label{fig:prep00FT}
\end{figure}

\begin{figure}[h]
	\centering
	\[
	\Qcircuit @C=.5em @R=.5em{
		\lstick{\mathrm{q_0}\rightarrow\ket{0}}&\qw     &\qw     &\qw      &\qw      &\qw&\push{\rule{0em}{1em}}\\ 
		\lstick{\mathrm{q_1}\rightarrow\ket{0}}&\qw     &\qw     &\qw      &\targ    &\qw&\push{\rule{0em}{1em}}\\
		\lstick{\mathrm{q_2}\rightarrow\ket{0}}&\qw     &\qw     &\targ    &\ctrl{-1}&\qw\\
		\lstick{\mathrm{q_3}\rightarrow\ket{0}}&\gate{H}&\ctrl{1}&\ctrl{-1}&\qw      &\qw\\
		\lstick{\mathrm{q_4}\rightarrow\ket{0}}&\qw     &\targ   &\qw      &\qw      &\qw
	}\]
	 \fcaption{A short but non-fault-tolerant circuit to prepare $\ket{00}_\mathrm{L}$.}
	\label{fig:prep00NFT}
\end{figure}

The flagging technique in \cite{Chao2017a} for the preparation of $\ket{00}_{\rm L}$ can be implemented with the given CNOT connectivity, see Fig.~\ref{fig:prep00FT}a.
We call this the fault-tolerant version one (FTv1).
The technique based on a circular layout proposed by \cite{Gottesman2016} cannot be implemented directly due to the connectivity.
A clever introduction of one SWAP gate permits to implement it still fault-tolerantly at the cost of 8 CNOT gates in total, see Fig.~\ref{fig:prep00FT}b. 
We call this the fault-tolerant version two (FTv2). We present below, the results for FTv1, which is more appropriate for the layout.
For both those circuits one can check that every possible single fault leads to a detectable error, or an error that stabilizes the prepared state.
We also tried a non-fault-tolerant one that has the advantage of being short, involving only 3 CNOTs, see \autoref{fig:prep00NFT}. 
We call this the non-fault-tolerant version (NFT).
For the preparation of $\ket{0+}_{\rm L}$, we want to create a Bell pair between $q_1$ and $q_3$ and between $q_2$ and $q_4$.
There is a missing connection between $q_1$ and $q_3$, but we can do a similar SWAP trick, see the resulting circuit in Fig.~\ref{fig:0PCatprep}a.
This circuit is not fault-tolerant as an undetected logical $Z_1\otimes Z_2 = Z\otimes\id\otimes\id\otimes Z$ can occur if the SWAP gate fails.
This is the only possible harmful logical error for this circuit.
We'll see below that we only use this circuit in cases were the possible undetected $Z_1\otimes Z_2$ error doesn't change the outcome of the sampling test.
The preparation for $\ket{00}_{\rm L}+\ket{11}_{\rm L}$ can itself be straightforwardly implemented fault-tolerantly, see Fig.~\ref{fig:0PCatprep}b.
To summarize, our sets of initial states and gates are
\begin{eqnarray*}
	S_{\rm FT} &=& \left \{\ket{00}, \ket{0+},\frac{\ket{00}+\ket{11}}{\sqrt{2}}\right \},\\
	G_{\rm FT} &=& \left \{X_1, X_2, Z_1, Z_2, H\otimes H\cdot\mathrm{SWAP}, \mathrm{C}Z\right \}.
\end{eqnarray*}

The $[[4,2,2]]$ code is only an error detecting code, that is, it can detect one Pauli error but cannot correct it.
This means that in place of error correction we have to rely on post-selection to remove errors.
In other words, we throw away runs where we detect that an error occurred, either from the ancilla measurement checking state preparation, or from the final measurement which indicates, when the parity of the outcomes is odd, that $S_Z$ has value $-1$.

As mentioned before, we cannot interleave rounds of error detection between state preparation and the gates.
That means that the final circuit can only be tolerant to a single fault during the whole computation, except for the specific undetectable $Z_1\otimes Z_2$ failure mentioned above and when we use the non-fault-tolerant version to prepare $\ket{00}_\mathrm{L}$.
\begin{figure}[h]
	\[
	\Qcircuit @C=.5em @R=.5em{
		\lstick{}&\qswap    &\qw&&\raisebox{-2.5em}{=}&&&\ctrl{1}&\gate{H}&\ctrl{1}&\gate{H}&\ctrl{1}&\qw\\
		\lstick{}&\qswap\qwx&\qw&&&&&\targ   &\gate{H}&\targ   &\gate{H}&\targ   &\qw\\
	}\]
	 \fcaption{Implementation of the SWAP gate.}
	\label{fig:SWAP}
\end{figure}
\begin{figure}[h]
	\centering
	\hspace{1cm}
	 \begin{minipage}{.45\textwidth}
		 \centering
		 {(a)}
		\[
		\Qcircuit @C=.5em @R=.5em{
			\lstick{\mathrm{q_1}\rightarrow\ket{0}}&\gate{H}&\qw      &\qswap    &\qw     &\push{\rule{0em}{1em}} \qw\\
			\lstick{\mathrm{q_2}\rightarrow\ket{0}}&\qw     &\targ    &\qswap\qwx&\ctrl{2}&\qw\\
			\lstick{\mathrm{q_3}\rightarrow\ket{0}}&\gate{H}&\ctrl{-1}&\qw       &\qw     &\qw\\
			\lstick{\mathrm{q_4}\rightarrow\ket{0}}&\qw     &\qw      &\qw       &\targ   &\qw
		}\]
	 \end{minipage}
	 \begin{minipage}{.45\textwidth}
		 \centering
		 {(b)}
		\[
		\Qcircuit @C=.5em @R=.5em{
			\lstick{\ket{0}}&\qw     &\targ    &\qw      \\
			\lstick{\ket{0}}&\gate{H}&\ctrl{-1}&\qw      \\
			\lstick{\ket{0}}&\gate{H}&\ctrl{1} &\qw      \\
			\lstick{\ket{0}}&\qw     &\targ    &\qw      
		}\]
	 \end{minipage}
	 \fcaption{(a) Preparation circuits of the logical state $\ket{0+}_{\rm L}$, with 5 CNOTs. (b) Preparation circuits of the logical Bell state $\left (\ket{00}_{\rm L}+\ket{11}_{\rm L}\right )/\sqrt{2}$ with 2 CNOTs.}\label{fig:0PCatprep}
\end{figure}

\subsection{Comments on the tested circuits}

Since we work with such a small system, we can exhaustively find and try all the logically equivalent circuits and optimize the bare version for each one.
Essentially, we are making it most likely for the bare version of the task to prevail.
From the set of states $S_{\rm FT}$, and the set of gates $G_{\rm FT}$, one can obtain 20 different stabilizer states.
All the states with their most efficient bare preparation circuit, using the native set of gates provided by IBM, are listed in \autoref{tab:circuits}.
A brute force approach was used to find them.

Note that adding more gates in the sequence would test states that are already tested with shorter circuits. 
Therefore those would certainly give worse results as we cannot interleave error detections between gates.
Hence we kept only these 20 different preparations.
This still extensively probes the computational capabilities of the $[[4,2,2]]$ code.

\begin{table}[p]
	\renewcommand{\arraystretch}{1.5}
	\centering
	\vspace{2cm}
	\begin{tabular}{c c c c c}
		\toprule
		Initial state & Unitary & Final state & \# Instructions \\\midrule
		$\ket{00}$ & $\id\otimes\id$ & $\ket{00}$ &5\\
		$\ket{0+}$ & $\id\otimes\id$ & $\frac{\ket{00}+\ket{01}}{\sqrt{2}}$ &6\\
		$\ket{00}$ & $\id\otimes X$ & $\ket{01}$ &6\\
		$\ket{00}$ & $X\otimes \id$ & $\ket{10}$ &6\\
		$\frac{\ket{00}+\ket{11}}{\sqrt{2}}$ & $\id\otimes\id$ & $\frac{\ket{00}+\ket{11}}{\sqrt{2}}$ &7\\
		$\ket{0+}$ & $\id\otimes Z$ & $\frac{\ket{00}-\ket{01}}{\sqrt{2}}$ &7\\
		$\ket{00}$ & $H\otimes H\cdot\mathrm{SWAP}$ & $\frac{\ket{00}+\ket{01}+\ket{10}+\ket{11}}{2}$ &7\\
		$\ket{0+}$ & $X\otimes \id$ & $\frac{\ket{10}+\ket{11}}{\sqrt{2}}$ &7\\
		$\ket{00}$ & $X\otimes X$ & $\ket{11}$ &7\\
		$\frac{\ket{00}+\ket{11}}{\sqrt{2}}$ & $\id\otimes Z$ & $\frac{\ket{00}-\ket{11}}{\sqrt{2}}$ &8\\
		$\frac{\ket{00}+\ket{11}}{\sqrt{2}}$ & $X\otimes \id$ & $\frac{\ket{10}+\ket{01}}{\sqrt{2}}$ &8\\
		$\ket{00}$ & $\id\otimes Z\cdot H\otimes H\cdot\mathrm{SWAP}$ & $\frac{\ket{00}-\ket{01}+\ket{10}-\ket{11}}{2}$ &8\\
		$\ket{00}$ & $Z\otimes\id\cdot H\otimes H\cdot\mathrm{SWAP}$ & $\frac{\ket{00}+\ket{01}-\ket{10}-\ket{11}}{2}$ &8\\
		$\ket{0+}$ & $X\otimes Z$ & $\frac{\ket{10}-\ket{11}}{\sqrt{2}}$ &8\\
		$\frac{\ket{00}+\ket{11}}{\sqrt{2}}$ & $\id\otimes ZX$ & $\frac{\ket{10}-\ket{01}}{\sqrt{2}}$ &9\\
		$\ket{00}$ & $Z\otimes Z\cdot H\otimes H\cdot\mathrm{SWAP}$ & $\frac{\ket{00}-\ket{01}-\ket{10}+\ket{11}}{2}$ &9\\
		$\ket{00}$ & $\mathrm{C}Z\cdot H\otimes H\cdot\mathrm{SWAP}$ & $\frac{\ket{00}+\ket{01}+\ket{10}-\ket{11}}{2}$ &10\\
		$\ket{00}$ & $\mathrm{C}Z\cdot \id\otimes Z\cdot H\otimes H\cdot\mathrm{SWAP}$ & $\frac{\ket{00}-\ket{01}+\ket{10}+\ket{11}}{2}$ &11\\
		$\ket{00}$ & $\mathrm{C}Z\cdot Z\otimes\id\cdot H\otimes H\cdot\mathrm{SWAP}$ & $\frac{\ket{00}+\ket{01}-\ket{10}+\ket{11}}{2}$ &11\\
		$\ket{00}$ & $\id\otimes X\cdot\mathrm{C}Z\cdot H\otimes H\cdot\mathrm{SWAP}\cdot X\otimes\id$ & $\frac{\ket{00}-\ket{01}-\ket{10}-\ket{11}}{2}$ &12\\\bottomrule
	\end{tabular}
	 \fcaption{
		The list of initial states and unitary circuits and the number of QASM instructions in the bare circuit (including final measurements). The final states are written exclusively in the computational basis since that is how they are measured.}
	\label{tab:circuits}
\end{table}

\section{Experimental results}
\label{sec:results}

The code and data can be found on GitHub \cite{Code}: it uses the Python SDK that can be found at \cite{SDK}.

\subsection{Parameters and runs}

Using the IBM chip one can run circuits in batches.
For each individual run, 8192 shots are done right one after another in a short time, where each shot outputs 5 bits as measurement outcomes.
We consider that the chip is exactly in the same conditions during each run.
We consider all the runs to be independent of one another but each to be done in different conditions, so sampling from different output distributions.

For each run, IBM also provides calibration data describing the state of the chip such as: gate error rates, readout error rates, $T_1$, $T_2$ and fridge temperature.
We give the average values that we observed for our runs in the appendix.

\subsection{Performance metric}

For each circuit run we want to compare the observed outcome distribution with the ideal one.
The ideal distribution is 8192 independent samples from a distribution with four possible outcomes occurring with probabilities $p_{00}$, $p_{01}$, $p_{10}$ and $p_{11}$.
The values for $p_{ij}$ can be read from \autoref{tab:circuits}.
Since we assume that the conditions stay identical during one run and that the 8192 shots are independent, we observe independent samples from a four-outcome distribution with some different probabilities $\tilde{p}_{00}$, $\tilde{p}_{01}$, $\tilde{p}_{10}$ and $\tilde{p}_{11}$.

We then use the statistical distance as a metric:
\[D = \frac{1}{2}\Big(\vert p_{00} - \tilde{p}_{00}\vert+\vert p_{01} - \tilde{p}_{01}\vert+\vert p_{10} - \tilde{p}_{10}\vert+\vert p_{11} - \tilde{p}_{11}\vert\Big).\]
This quantity is estimated for each run by
\[\hat{D} = \frac{1}{2}\left (\left \vert p_{00} - \frac{n_{00}}{n_{\rm valid}}\right \vert+\left \vert p_{01} - \frac{n_{01}}{n_{\rm valid}}\right \vert+\left \vert p_{10} - \frac{n_{10}}{n_{\rm valid}}\right \vert+\left \vert p_{11} - \frac{n_{11}}{n_{\rm valid}}\right \vert\right ),\]
where $n_{ij}$ is the number of observation of outcome $ij$ after post-selection and $n_{\rm valid}$ is the number of shots kept after post-selection.
This estimator for $D$ is slightly biased except for the case where only one of the $p_{ij}$ is non-zero (because in this case it becomes linear).
We use this estimator to keep the analysis simple.

Each of the runs has some different $\tilde{p}_{ij}$ and we have no information about how the $\tilde{p}_{ij}$s vary.
Therefore we will assume that there are fluctuations around the mean of $\hat{D}$ following some unknown normal distribution and use this model to compute confidence intervals \cite{Walpole1985}.
This means that the final data points and their confidence intervals don't exactly reflect knowledge of $D$ but only of $\hat{D}$ which we believe is still a valid quantity to characterize the performance of the circuits.

\subsection{Comparisons}

\begin{figure}[h]
	\centering
	 \fcaption{Comparison of the different pairs of qubits to implement the bare circuits together with the number of instructions for each circuit.}\label{fig:bareresults}
	\includestandalone[scale=1.3,mode=buildnew]{bare_stat_dists}
\end{figure}

\begin{figure}[h]
	\centering
	 \fcaption{Comparing encoded performance with bare qubit pair $[2-0]$.
		The performance is defined as the statistical distance to the ideal outcome distribution.
		The figure shows the difference between encoded and bare performance.
		The error bars show confidence intervals at $99\%$.}\label{fig:encodedresults}
	\includestandalone[scale=1.3,mode=buildnew]{enc_stat_dist_diff_20}
\end{figure}

\begin{table}[h]
	\centering
	\begin{filecontents*}{AveragePerf.dat}
class averagePerf PostSelectR
Bare[1-0] 0.0672178879495609   1.0 
Bare[2-0] 0.055036253140751126 1.0 
Bare[2-1] 0.0648934692934251   1.0 
Bare[2-4] 0.06984813141108422  1.0 
Bare[3-2] 0.06273241831836669  1.0 
Bare[3-4] 0.07725998894284998  1.0 
FTv1 0.045068083100526755 0.6467531826829663
NFT 0.060496868329305044  0.7084630103444985
\end{filecontents*}
\pgfplotstabletypeset[columns/class/.style={string type, column name={Implementation}},
	columns/PostSelectR/.style={multiply by=1, fixed zerofill, column name={Post-selection ratio}, precision=2},
	columns/averagePerf/.style={multiply by=100, fixed zerofill, column name={Avg.Perf.($\times10^{-2}$)}, precision=2}, every head row/.style={before row=\toprule,after row=\midrule},
	every last row/.style={after row=\bottomrule}]{AveragePerf.dat}
	 \tcaption{Average performance for the different bare and encoded possible implementations with the post-selection ratio (ratio of data kept). The encoded implementations only differ in how they prepare the state $\ket{00}_\mathrm{L}$.}\label{tab:avgperf}
\end{table}

We first need to decide on what pair of bare qubits is the best to be compared to the encoded qubits.
It is not immediately clear how to choose this given only the calibration numbers.
Thus we tried out the six different connected pairs and we found the performance shown in \autoref{fig:bareresults}.
There is no clear ``best pair'' but one can see that the pairs $[2-0]$ or $[3-2]$ seem to be slightly better.
Since we don't have a systematic method for predicting the best pair given the calibration data and the circuit, we will look at average performance for each pair over all the circuits.
When averaging, we find the pair $[2-0]$ to be the best, see \autoref{tab:avgperf}.
The second observation that we can make based on \autoref{fig:bareresults} is that the number of instructions does not explain the performance.
It seems that the type of state sampled from is more important.
Roughly it is easier to sample from equal superposition of the four computational basis states, than from equal superposition of two, than from just one.
This can be understood with the fact that with more states in equal superposition there are less Pauli errors or readout errors that can affect the outcome distribution.

We then go on, comparing the encoded versions of the circuits to the elected best pair in \autoref{fig:encodedresults}.
The encoded circuits using the preparations $\ket{0+}_\mathrm{L}$ and $\ket{00}_\mathrm{L}+\ket{11}_\mathrm{L}$ are short and fault-tolerant and indeed show better performance than the bare ones.
For the different preparations for $\ket{00}_\mathrm{L}$, the fault-tolerant version FTv1 is, except in a few cases, better than the non-fault-tolerant one despite being substantially longer.
This shows that fault-tolerant design of circuits can be useful.
Although they both compare unfavourably to the bare version, except for 4 circuits for FTv1 and only 1 circuit for NFT.

We cannot make an absolute statement as for some circuits the bare version is always better.
We also don't want to cherry pick the best version for each circuit since we don't have a systematic way of predicting the best version.
Average performance when fixing a preparation for $\ket{00}_\mathrm{L}$ are shown in \autoref{tab:avgperf}.
One can see that FTv1, with $4.51\times 10^{-2}$, is better than the best bare implementation with $5.5\times 10^{-2}$, whereas NFT ($6.05\times 10^{-2}$) is worse.

\section{Conclusion}

In conclusion, we have shown that already on the IBM~5Q chip one can improve some quantum computation task, namely sampling from a class of states, by using error detection and fault-tolerant design of circuits.
This improvement is only on average over 20 different states that the $[[4,2,2]]$ code can fault-tolerantly prepare.
We also can see that shorter but non fault-tolerant circuits can be bested by longer but fault-tolerant circuits, showing the usefulness of these.
As better and better hardware is developed, with more physical qubits, more connectivity and smaller error rates, demonstrations of fault-tolerance will become easier to produce and become more convincing.
The set of gates shown to be fault-tolerant in this paper is very restricted.
For the $[[4,2,2]]$ code a few more qubits and connections would be needed to realize fully fault-tolerant circuits with error detection in between logical units.
Being able to demonstrate the fault-tolerance of larger gate sets, for example the whole Clifford group for several logical qubits would be an important milestone towards harnessing universal quantum computation.\\

\nonumsection{Acknowledgements}
\noindent
I would like to thank the IBM Quantum Experience team for providing extensive access to the chip and support on how to interface with it.
I also would like to thank Barbara Terhal for valuable comments and feedback.
CV acknowledge support through the EU via the ERC GRANT EQEC and support by the Excellence Initiative of DFG.
The views expressed are those of the author and do not reflect the official policy or position of IBM or the IBM Quantum Experience team.

\bibliographystyle{hunsrtnat}
\bibliography{article_fault_tolerance_demonstration}
\newpage
\appendix
\noindent
This appendix presents the calibration data for the experiments, averaged over all the different runs as well as the observed standard deviation. 
It also shows previous runs on a previous iteration of the chip, called Sparrow, which was showing better performance but has been taken down by IBM.
The current iteration is named Raven: the more extensive runs presented in this paper were done on this chip.
Note that for the runs on Sparrow, there were only 4 different calibrations and 36 runs for each circuit.
For Raven we have more than 100 runs for each circuit and almost a new calibration per run.

\pgfplotstableset{
	columns/name/.style={string type, column name=qubit},
	columns/T1/.style={multiply by=1000000, fixed zerofill, column name=$T_1 (\si{\micro\second})$, precision=0},
	columns/T2/.style={multiply by=1000000, fixed zerofill, column name=$T_2 (\si{\micro\second})$, precision=0},
	columns/gateError/.style={multiply by=100, fixed, column name={gate error (\%)}, precision=2},
	columns/readoutError/.style={multiply by=100, fixed zerofill, column name={readout error (\%)}, precision=1},
	columns/sigma(T1)/.style={multiply by=1000000, fixed zerofill, column name=$\sigma[T_1]$, precision=1},
	columns/sigma(T2)/.style={multiply by=1000000, fixed zerofill, column name=$\sigma[T_2]$, precision=1},
	columns/sigma(gateError)/.style={multiply by=100, fixed, column name={$\sigma[\textrm{gate error}]$}, precision=3},
	columns/sigma(readoutError)/.style={multiply by=100, fixed, column name={$\sigma[\textrm{readout error}]$}, precision=2},
	columns/qubits/.style={string type, column name={pair}},
	every head row/.style={before row=\toprule,after row=\midrule},
	every last row/.style={after row=\bottomrule}
}

\begin{table}[h]
	\centering
	\begin{tabular}{c c c}
		\toprule
		chip & Temperature (\si{\milli\kelvin}) \\
		\midrule
		Raven & 21 \\
		Sparrow & 19\\
		\bottomrule
	\end{tabular}
	 \tcaption{Average fridge temperatures for the two chips at the time of the runs in \si{\milli\kelvin}.}
	\label{tab:temperature}
\end{table}

\begin{table}[h]
	\centering
	\begin{minipage}{.45\textwidth}
		\centering
		 {(a) Raven}
		\begin{filecontents*}{single_q.dat}
name T1 sigma(T1) T2 sigma(T2) gateError sigma(gateError) readoutError sigma(readoutError)
Q0 4.5761394992335204e-05 6.533104754402174e-06 3.982845828162639e-05 7.2808585505352315e-06 0.0010241296792490096 0.00020894372540194745 0.04111019052485583 0.00629192826642451
Q1 5.679194101759252e-05 6.84146555971259e-06 4.473533469596321e-05 6.335786521132446e-06 0.0008130544878602845 0.00034692020529052754 0.047309000656982264 0.009594421583335757
Q2 3.900369370027009e-05 5.073455415762868e-06 4.0379294109059055e-05 3.1968505435994042e-06 0.0015285727273005652 0.0002353252803414579 0.02727783049857654 0.0035998379461931705
Q3 4.003238557558946e-05 4.803942221847981e-06 5.741640630702971e-05 1.2479412948549455e-05 0.0014878815292376695 0.0001573662914243094 0.062006387327542156 0.011051925084748905
Q4 5.3769647419519674e-05 9.146820619050877e-06 2.0946481494999635e-05 1.3474556006918988e-06 0.001855114961678993 0.0002606179176798011 0.0510334695963209 0.01103319725666919
\end{filecontents*}
\pgfplotstabletypeset[columns={name, T1, sigma(T1), T2, sigma(T2)}]{single_q.dat}
	\end{minipage}
	\qquad
	\begin{minipage}{.45\textwidth}
		\centering
		 {(b) Sparrow}
		\begin{filecontents*}{sparrow_single_q.dat}
name T1 sigma(T1) T2 sigma(T2) gateError sigma(gateError) readoutError sigma(readoutError)
Q0 48.59444444444444 8.90339961427 39.40833333333333 4.4625213974  0.2297222222222222  0.00897097514553 1.883333333333333  0.215380799722
Q1 46.46111111111111 11.0131557133 38.97222222222222 13.3308875072 0.29888888888888887 0.0986514002203  8.255555555555555  2.19449367033
Q2 55.53611111111111 1.65800294494 90.68888888888888 9.85443435937 0.4227777777777777  0.0851016095654  1.4555555555555557 0.289102485139
Q3 47.45833333333333 4.85342805539 62.19999999999999 12.0036799913 0.5152777777777777  0.212541317697   12.283333333333333 3.60389449839
Q4 55.31666666666666 5.99761526683 80.67777777777778 13.1859797409 0.33305555555555555 0.0271640534985  6.55               0.831163842653
\end{filecontents*}
\pgfplotstabletypeset[columns/T1/.style={multiply by=1, fixed zerofill, column name=$T_1 (\si{\micro\second})$, precision=0},
		columns/T2/.style={multiply by=1, fixed zerofill, column name=$T_2 (\si{\micro\second})$, precision=0},
		columns/sigma(T1)/.style={multiply by=1, fixed zerofill, column name=$\sigma[T_1]$, precision=1},
		columns/sigma(T2)/.style={multiply by=1, fixed zerofill, column name=$\sigma[T_2]$, precision=1},
		columns={name, T1, sigma(T1), T2, sigma(T2)},
		]{sparrow_single_q.dat}
	\end{minipage}
	 \tcaption{Average $T_1$ and $T_2$ parameters for the runs together with their standard deviation all in \si{\micro\second}.}
	\label{tab:T1T2}
\end{table}

The two chips are similar, the gate and readout error rates have been slightly improved with Raven but the fridge temperature as well as $T_1$ and $T_2$ times became a bit worse.
The layout differs only in the orientation of the CNOTs, see \autoref{fig:IBMchipSparrow}.
The circuits run on Sparrow are showed in \autoref{fig:prep00Sparrow} and \autoref{fig:0PCatprepSparrow}.
Comparison of the bare and encoded performance between Raven and Sparrow are presented in \autoref{fig:BareRavenvsSparrow} and \autoref{fig:EncodedRavenvsSparrow}.
One can see that the previous iteration was performing better.

\begin{table}[h]
	
	\centering
	\begin{minipage}{\textwidth} 
		\centering	
		 {(a) Raven}\\
		\pgfplotstabletypeset[columns={name, gateError, sigma(gateError), readoutError, sigma(readoutError)}]{single_q.dat}
	\end{minipage}\vspace{1em}
	
	\begin{minipage}{\textwidth}
		\centering	  
		 {(b) Sparrow}\\
		\pgfplotstabletypeset[columns/gateError/.style={multiply by=1, fixed zerofill, column name={gate error (\%)}, precision=2},
		columns/readoutError/.style={multiply by=1, fixed zerofill, column name={readout error (\%)}, precision=1},
		columns/sigma(gateError)/.style={multiply by=1, fixed, column name={$\sigma[\textrm{gate error}]$}, precision=3},
		columns/sigma(readoutError)/.style={multiply by=1, fixed zerofill, column name={$\sigma[\textrm{readout error}]$}, precision=2},
		columns={name, gateError, sigma(gateError), readoutError, sigma(readoutError)}]{sparrow_single_q.dat}
	\end{minipage}
	 \tcaption{Average single-qubit gate and readout error rates for the runs together with their standard deviation, all in percent.}
	\label{tab:singleError}
\end{table}

\begin{table}[h]
	\centering
	\begin{minipage}{.45\textwidth}
		\centering
		 {(a) Raven}
		\begin{filecontents*}{multi_q.dat}
qubits gateError sigma(gateError)
1-0 0.02231822679352285 0.003190640563855979
2-0 0.02289358657039413 0.0025336928141787124
2-1 0.02658698645701568 0.003597721690675582
2-4 0.03932712314052565 0.007755219128397571
3-2 0.021570207415827828 0.002970777536571876
3-4 0.027076798505243874 0.003767389492659444
\end{filecontents*}
\pgfplotstabletypeset[
		columns/gateError/.style={multiply by=100, fixed zerofill, column name={gate error (\%)}, precision=1},
		columns/sigma(gateError)/.style={multiply by=100, fixed zerofill, column name={$\sigma[\textrm{gate error}]$}, precision=2}]{multi_q.dat}
	\end{minipage}
	\qquad
	\begin{minipage}{.45\textwidth}
		\centering
		 {(b) Sparrow}
		\begin{filecontents*}{sparrow_multi_q.dat}
qubits gateError sigma(gateError)
1-0 4.012499999999999  0.73841186565
2-0 2.9780555555555552 0.237671802708
2-1 4.599444444444444  0.863169046288
2-4 3.6783333333333337 0.335182371586
3-2 6.209444444444444  0.8964341224
3-4 5.756111111111111  2.0538504189
\end{filecontents*}
\pgfplotstabletypeset[
		columns/gateError/.style={multiply by=1, fixed zerofill, column name={CNOT error (\%)}, precision=1},
		columns/sigma(gateError)/.style={multiply by=1, fixed zerofill, column name={$\sigma[\textrm{CNOT error}]$}, precision=2}]{sparrow_multi_q.dat}
	\end{minipage}
	 \tcaption{Average CNOT gate error rates and their standard deviation for the runs, all in percent.}
	\label{tab:multiError}
\end{table}

\begin{figure}[h]
	\centering
	\begin{tikzpicture}[node distance = 5em]
	\draw
	node[draw, circle,label = above:{\co{red}4}] (q2) {$q_2$}
	node[draw, circle, below right of = q2, label = right:{\co{red}1}] (q1) {$q_1$}
	node[draw, circle, below left of = q2, label = left:{\co{red}2}] (q3) {$q_3$}
	node[draw, circle, above left of = q2, label = left:{\co{red}3}] (q4) {$q_4$}
	node[draw, circle, above right of = q2, label = right:{\co{red}A}] (q0) {$q_0$};
	
	\path [draw, vecArrow] (q0) -- (q2)
	node[pos=0.5, below] (p) {};
	
	\path [draw, vecArrow] (q1) -- (q2)
	node[pos=0.5, below] (q) {};
	
	\path [draw, vecArrow] (q3) -- (q2)
	node[pos=0.5, below] (r) {};
	
	\path [draw, vecArrow] (q4) -- (q2)
	node[pos=0.5, below] (s) {};
	
	\path [draw, vecArrow] (q3) -- (q4)
	node[pos=0.5, below] (r) {};
	
	\path [draw, vecArrow] (q0) -- (q1)
	node[pos=0.5, below] (s) {};
	\end{tikzpicture}
	 \fcaption{Layout of the Sparrow chip.}\label{fig:IBMchipSparrow}
\end{figure}

\begin{figure}[h]
	\centering
	\[
	\Qcircuit @C=.5em @R=.5em{
		\lstick{\mathrm{q_0}\rightarrow\ket{0}}&\qw     &\qw     &\qw      &\qw       &\qw      &\gate{H}&\ctrl{2}&\ctrl{1}&\gate{H}&\meter&\cw\\ 
		\lstick{\mathrm{q_1}\rightarrow\ket{0}}&\qw     &\qw     &\qw      &\qswap    &\qw      &\gate{H}&\qw     &\targ   &\gate{H}&\qw   &\qw\\
		\lstick{\mathrm{q_2}\rightarrow\ket{0}}&\qw     &\qw     &\targ    &\qswap\qwx&\targ    &\gate{H}&\targ   &\qw     &\gate{H}&\qw   &\qw\\
		\lstick{\mathrm{q_3}\rightarrow\ket{0}}&\gate{H}&\ctrl{1}&\qw      &\qw       &\ctrl{-1}&\qw     &\qw     &\qw     &\qw     &\qw   &\qw\\
		\lstick{\mathrm{q_4}\rightarrow\ket{0}}&\qw     &\targ   &\ctrl{-2}&\qw       &\qw      &\qw     &\qw     &\qw     &\qw     &\qw   &\qw
	}\]
	 \fcaption{
		The circuit previously implemented on the IBM~5Q Sparrow chip preparing the logical state $\ket{00}_{\rm L}$.
		The SWAP gate is implemented via the circuit in \autoref{fig:SWAP}.
		If the SWAP gate fails, it can introduce Pauli $X$ errors on $q_1$ and $q_2$, which would not be detected and which constitute a logical $X_1\otimes X_2$ error.}
	\label{fig:prep00Sparrow}
\end{figure}

\begin{figure}[h]
	\centering
	 \begin{minipage}{.18\textwidth}
		 \centering
		 {(a)}
		\[
		\Qcircuit @C=.5em @R=.5em{
			\lstick{\mathrm{q_1}\ket{0}}&\qw     &\qw      &\qswap    &\qw      &\push{\rule{0em}{1em}} \qw\\
			\lstick{\mathrm{q_2}\ket{0}}&\qw     &\targ    &\qswap\qwx&\targ    &\qw\\
			\lstick{\mathrm{q_3}\ket{0}}&\gate{H}&\ctrl{-1}&\qw       &\qw      &\qw\\
			\lstick{\mathrm{q_4}\ket{0}}&\gate{H}&\qw      &\qw       &\ctrl{-2}&\qw
		}\]
	 \end{minipage}
	\qquad\qquad
	 \begin{minipage}{.18\textwidth}
		 \centering
		 {(b)}
		\[
		\Qcircuit @C=.5em @R=.5em{
			\lstick{\ket{0}}&\gate{H}&\ctrl{1}&\qw      \\
			\lstick{\ket{0}}&\qw     &\targ   &\qw      \\
			\lstick{\ket{0}}&\gate{H}&\ctrl{1}&\qw      \\
			\lstick{\ket{0}}&\qw     &\targ   &\qw      
		}\]
	 \end{minipage}
	 \fcaption{(a) Preparation circuits of the logical state $\ket{0+}_{\rm L}$, with 5 CNOTs. 
		If the SWAP gate fails, it can introduce Pauli $X$ errors on $q_1$ and $q_2$, which would not be detected and which constitute a logical $X_1\otimes X_2$ error.
		(b) Preparation circuits of the logical Bell state $\left (\ket{00}_{\rm L}+\ket{11}_{\rm L}\right )/\sqrt{2}$ with 2 CNOTs.}\label{fig:0PCatprepSparrow}
\end{figure}

\begin{figure}[p]
	\vspace{3cm}
	 \fcaption{Comparison of the performances of the bare circuits between the Raven and Sparrow chips. Except for the four circuits involving a two-qubit gate, Sparrow was showing better performance. This is probably due to improvement of the two-qubit gates but shorter $T_1$ and $T_2$ times on Raven, see \autoref{tab:T1T2} and \autoref{tab:multiError}.}
	\begin{center}
		\includestandalone[scale=1.3,mode=buildnew]{compare_raven_sparrow_bare}
	\end{center}
	\label{fig:BareRavenvsSparrow}
\end{figure}

\begin{figure}[p]
	\vspace{2cm}
	 \fcaption{Comparing encoded versions on Raven and Sparrow.
		The $\ket{00}_\mathrm{L}$ and $\ket{0+}_\mathrm{L}$ preparations were suffering from possible $X_1X_2$ logical errors.}
	\begin{center}
		\includestandalone[scale=1.3,mode=buildnew]{compare_raven_sparrow_encoded}
	\end{center}
	\label{fig:EncodedRavenvsSparrow}
\end{figure}

\end{document}

%% file: bare_stat_dists.tex
\begin{tikzpicture}

	\begin{axis}[
			axis y line=right,
			axis x line=none,
			name=qasm count,
			ylabel=\#instructions,
			y label style={font=\tiny, xshift=-1.1cm, yshift=.8cm},
			ytick distance=2,
			minor y tick num=1,
			ymajorgrids,
			yminorgrids,
			ymax=12.5,
			ymin=4,
			ytick align=inside,
			y tick label style={font=\tiny, yshift=.1cm, xshift=.1cm, anchor=east}
			]

		\addplot[
			color=black!60,
			mark=none,
			]
		table [x = index, y = qasm_count]{bare_qasm_count.dat};

	\end{axis}

    \begin{axis}[
    	name=absolute,
        ylabel=Average statistical error,
        ylabel style={font=\tiny},
        xtick=data,
        xticklabels from table={circuit_names.dat}{name},
        x tick label style={font=\tiny, rotate=70, anchor=east},
        y tick label style={font=\tiny},
        scaled ticks=false, 
        tick label style={/pgf/number format/fixed},
        ytick distance=.02,
        ymin=0,
        ymax=.17,
        legend pos=outer north east,
        legend style={font=\tiny},
        legend entries={
        		pair 1-0,
        		pair 2-0,
        		pair 2-1,
        		pair 2-4,
        		pair 3-2,
        		pair 3-4,
        		\#instr.
        		},
        legend cell align=left
        ]

		\addlegendimage{only marks, mark = x, color1}
		\addlegendimage{only marks, mark = +, color2}
		\addlegendimage{only marks, mark = star, color3}
		\addlegendimage{only marks, mark = triangle, color4}
		\addlegendimage{only marks, mark = o, color5}
		\addlegendimage{only marks, mark = diamond, color6}
		\addlegendimage{only marks, mark = -, black!60}

		
        \addplot[
            color=color1,
            mark=x,
            only marks,
        ]
        table [x = index, y = stat_dist, y error = conf_int99]{bare10.dat};

        \addplot[
			color=color2,
			mark=+,
			only marks,
			]
		table [x = index, y = stat_dist, y error = conf_int99]{bare20.dat};
		
        \addplot[
			color=color3,
			mark=star,
			only marks,
			]
		table [x = index, y = stat_dist, y error = conf_int99]{bare21.dat};

        \addplot[
			color=color4,
			mark=triangle,
			only marks,
			]
		table [x = index, y = stat_dist, y error = conf_int99]{bare24.dat};

        \addplot[
			color=color5,
			mark=o,
			only marks,
			]
		table [x = index, y = stat_dist, y error = conf_int99]{bare32.dat};

        \addplot[
			color=color6,
			mark=diamond,
			only marks,
			]
		table [x = index, y = stat_dist, y error = conf_int99]{bare34.dat};

    \end{axis}

\end{tikzpicture}

%% file: enc_stat_dist_diff_20.tex
\begin{tikzpicture}

    \begin{axis}[
  			  	name=twozero,
  			  	ylabel=Differences of performance (Encoded - Bare), 
  			  	ylabel style={font=\tiny, yshift=-.3cm},    
       	 		xtick={1,2,3,4,5,6,7,8,9,10,11,12,13,14,15,16,17,18,19,20},
       	 		xticklabels from table={circuit_names.dat}{name},
       	 		x tick label style={font=\tiny, rotate=70, anchor=east}, 	
       	 		y tick label style={font=\tiny},
       	 		minor y tick num=1,
       	 		xmin=-1,
       	 		xmax=22,
       	 		ymin=-.1,
       	 		ymax=.1,
  	     		ymajorgrids,
  	     		yminorgrids,
  	     		legend pos=outer north east,
  	     		legend style={font=\tiny},
  	     		legend entries={
  	     			$\ket{00}$FTv1,
  	     			$\ket{00}$NFT,
  	     			$\ket{0+}$,
  	     			$\ket{00}+\ket{11}$
  	     		},
  	     		legend cell align=left
     	  		]
     	  		
     	  		\addlegendimage{only marks, mark = +, color2}
     	  		\addlegendimage{only marks, mark = star, color6}
     	  		\addlegendimage{only marks, mark = triangle, color3}
     	  		\addlegendimage{only marks, mark = diamond, color5}
     	  		
		\addplot [draw=black,style=thick] coordinates {(-2,0) (22,0)}; 
		\addplot[
				color=color2,
				mark=+,
				only marks,
				error bars/.cd, 
				y dir=both,
				y explicit,
				error bar style={line width=0.15pt},
				    			error mark options={
				     	 		rotate=90,
				     	 		line width=.15pt,
				     	 		mark size=1pt}
				]
			table [x = index, y = stat_dist_diff, y error = conf_int99]{encoded00ftv1-bare20.dat};
			
			
		\addplot[
				color=color6,
				mark=star,
				only marks,
				error bars/.cd, 
				y dir=both,
				y explicit,
				error bar style={line width=0.15pt},
				error mark options={
					rotate=90,
					line width=.15pt,
					mark size=1pt}
				]
			table [x = index, y = stat_dist_diff, y error = conf_int99]{encoded00nftv1-bare20.dat};

		\addplot[
				color=color3,
				mark=triangle,
				only marks,
				error bars/.cd, 
				y dir=both,
				y explicit,
				error bar style={line width=0.15pt},
				error mark options={
					rotate=90,
					line width=.15pt,
					mark size=1pt}
				]
			table [x = index, y = stat_dist_diff, y error = conf_int99]{encoded0+-bare20.dat};

		\addplot[
				color=color5,
				mark=diamond,
				only marks,
				error bars/.cd, 
				y dir=both,
				y explicit,
				error bar style={line width=0.15pt},
				error mark options={
					rotate=90,
					line width=.15pt,
					mark size=1pt}
				]
			table [x = index, y = stat_dist_diff, y error = conf_int99]{encoded00+11-bare20.dat};

\end{axis}

\end{tikzpicture}

%% file: compare_raven_sparrow_bare.tex
\begin{tikzpicture}
    \begin{axis}[
    			name=absolute,
    			ylabel=Average statistical error,
    			ylabel style={font=\tiny},
        		xtick={1,2,3,4,5,6,7,8,9,10,11,12,13,14,15,16,17,18,19,20},
        		xticklabels from table={circuit_names.dat}{name},
        		x tick label style={font=\tiny, rotate=70, anchor=east},
        		y tick label style={font=\tiny},
        		scaled ticks=false, 
        		tick label style={/pgf/number format/fixed},
        		ytick distance=.02,
        		ymin=-0.01,
        		ymax=.17,
        		ymajorgrids,
        		yminorgrids,
				legend entries={
                		Sparrow 2-0,
                		Raven 2-0
                		},
        		legend style={font=\scriptsize},
        		legend pos = north east,
        		legend cell align=left
        		]

		\addlegendimage{only marks, mark = x, color6}
		\addlegendimage{only marks, mark = star, color2}

		\addplot[
				color=color6,
				mark=x,
				mark size=3,
				only marks,
				]
			table [x = index, y = stat_dist, y error = conf_int99]{sparrow_bare20.dat};

        \addplot[
            	color=color2,
            	mark=star,
            	mark size=3,
            	only marks,
        		]
        	table [x = index, y = stat_dist, y error = conf_int99]{bare32.dat};

    \end{axis}
    
\end{tikzpicture}

%% file: compare_raven_sparrow_encoded.tex
\begin{tikzpicture}
    \begin{axis}[
    			name=absolute,
    			ylabel=Average statistical error,
    			ylabel style={font=\tiny},
        		xtick={1,2,3,4,5,6,7,8,9,10,11,12,13,14,15,16,17,18,19,20},
        		xticklabels from table={circuit_names.dat}{name},
        		x tick label style={font=\tiny, rotate=70, anchor=east},
        		y tick label style={font=\tiny},
        		scaled ticks=false, 
        		tick label style={/pgf/number format/fixed},
        		ytick distance=.01,
        		ymin=-0.0,
        		ymax=.15,
        		ymajorgrids,
        		yminorgrids,
				legend entries={
                		Sparrow,
                		Raven FTv1,
                		Raven NFT,
                		Raven $\ket{0+}$,
                		Raven $\ket{00}+\ket{11}$
                		},
        		legend style={font=\tiny},
        		legend pos = outer north east,
        		legend cell align=left
        		]

		\addlegendimage{only marks, mark = o, black}
		\addlegendimage{only marks, mark = +, color2}
		\addlegendimage{only marks, mark = star, color6}
		\addlegendimage{only marks, mark = triangle, color3}
		\addlegendimage{only marks, mark = diamond, color5}

		\addplot[
				color=black,
				mark=o,
				mark size=3,
				only marks,
				]
			table [x = index, y = stat_dist, y error = conf_int99]{sparrow_encoded.dat};

        
        \addplot[
        color=color2,
        mark=+,
        mark size=3,
        only marks,
        ]
        table [x = index, y = stat_dist, y error = conf_int99]{encoded00ftv1.dat};
        
        \addplot[
        color=color6,
        mark=star,
        mark size=3,
        only marks,
        ]
        table [x = index, y = stat_dist, y error = conf_int99]{encoded00nftv1.dat};
        
        \addplot[
        	color=color3,
        	mark=triangle,
        	mark size=3,
        	only marks,
        	]
        	table [x = index, y = stat_dist, y error = conf_int99]{encoded0+.dat};
        \addplot[
        	color=color5,
        	mark=diamond,
        	mark size=3,
        	only marks,
        	]
        	table [x = index, y = stat_dist, y error = conf_int99]{encoded00+11.dat};

    \end{axis}
    
\end{tikzpicture}